\documentclass{elsart}

\usepackage{graphicx}
\usepackage[english]{babel}

\begin{document}

\begin{frontmatter}

\title{Hidden invariance in Gurzadyan-Xue cosmological models}

\author{G.V. Vereshchagin \and G. Yegorian}
\ead{veresh@icra.it}
\ead{gegham@icra.it}

\address{ICRANet, P.le della Repubblica 10, 65100 Pescara, Italy and \\
ICRA, Dip. Fisica, Univ. ``La Sapienza'', P.le A. Moro 5, 00185 Rome, Italy}


\begin{abstract}
The dark energy formula derived by Gurzadyan and Xue which leads to a value fitting the SN data, provides a scaling relation between physical constants and cosmological parameters and defines a set of cosmological models. In previous works we considered several of those models and derived cosmological equations for each of them. In this letter we present the phase portrait analysis of those models. Surprisingly, we found that the separatrix in the phase space which determines the character of solutions depends solely on the value of the current matter density, namely: at $\Omega_m>2/3$ the equations describe Friedmannian Universe with the classical singularity at the beginning, while at $\Omega_m<2/3$ all solutions for the considered models start with zero density and non-vanishing scale factor. Even more remarkable, the value $\Omega_{sep}=2/3$, which defines the separatrix, is the same for all models. The latter reveales an underlying invariance hidden in the models, possibly, due to the basic nature of the GX-scaling.
\end{abstract}

\end{frontmatter}

Among the variety of models for dark energy proposed in the literature a remarkable fit to the SN data is given by the Gurzadyan-Xue formula \cite{GX} for the vacuum fluctuations energy density. It is shown that, if one takes into account not all but only relevant modes for vacuum fluctuations, then one arrives to
\begin{eqnarray}
	\rho_{GX}=\frac{\pi}{8}\,\frac{c^4}{G}\,\frac{1}{a^2},
\label{Lambda}
\end{eqnarray}
where $c$ is the speed of light, $G$ is the gravitational constant and $a$ is the scale factor of the Universe. This formula is derived when only $l=0$ modes of vacuum fluctuations are taken into account. In \cite{GX} the authors state that this can be a representation of the homogeneity and isotropy of the Universe, i.e. of the FRLW metric.

Taking the Hubble length for the value of the scale factor today, we find the following value of the density parameter for dark energy\footnote{A factor 1/2 was missing in our previous papers. The value of $\Omega_\Lambda$ obtained here is even closer to the observational one.}
\begin{eqnarray}
	\Omega_\Lambda\simeq 3.29.
\end{eqnarray}

Formula (\ref{Lambda}) opens interesting possibilities related to the scaling between physical constants and the scale of the Universe leading to the idea of the world constants variation, actively discussed in the recent literature (see e.g. \cite{barrow}). In previous papers \cite{Ver05,Ver06} we derived cosmological equations assuming variation of the physical constants such as the speed of light and the gravitational constant. Certain observational aspects of the GX-formula (1) are discussed in \cite{DG}.\footnote{Recently a similar scaling is given in \cite{P}, based on dimensionality considerations (i.e. with an accuracy of an unknown numerical factor), as a geometric mean of infrared and ultraviolet scales. However, it is not clear how such geometric mean can be self-consistently derived via conventional field theory methods.} 

In this Letter we present phase portrait analysis (cf. \cite{Bel85}) of these cosmological equations. Surprisingly, we find that, although equations look very differently, the character of solutions for the two pairs of models is similar. At the same time, even more striking feature of the models came out: the position of separatrix in the phase space of dynamical variables is determined by a single quantity, the matter density parameter $\Omega_m$, moreover, its value is the same for all models! This invariance should be connected to the scaling (\ref{Lambda}) between world constants and the scale factor of the Universe.

Following \cite{Ver06}, we consider 4 models and present equations for them:

{\bf Model I.}
If neither the speed of light nor the gravitational constant vary with time, then the cosmological constant scales like $\Lambda\propto a^{-2}$, as can be seen from (\ref{Lambda}). The cosmological equations for this model are the following
\begin{equation}
\begin{array}{l}
\displaystyle{\left(\frac{da}{adt}\right)^2+\frac{c^2}{a^2}\left(k-\frac{2\pi^2}{3}\right)=\frac{8\pi G}{3}\mu}, \\
\displaystyle{\dot\mu+3\left(\frac{da}{adt}\right)\mu=\frac{\pi}{2G}\left(\frac{c}{a}\right)^2\left(\frac{da}{adt}\right)},
\end{array}
\end{equation}
where $\mu$ is the matter density, $k$ is the spatial curvature.

{\bf Model II.}
To keep interpretation of $\Lambda=\frac{8\pi G}{c^2}\rho_{GX}$ as a cosmological constant, we require the varying speed of light, $c=\left(\frac{\Lambda}{2\pi^2}\right)^{1/2}a$, but $G$=const. Then the cosmological equations are reduced to the following system
\begin{equation}
\begin{array}{l}
\displaystyle{\left(\frac{da}{adt}\right)^2-\frac{\Lambda}{3}\left(1-\frac{3k}{2\pi^2}\right)=\frac{8\pi G}{3}\mu}, \\
\displaystyle{\dot\mu+3\left(\frac{da}{adt}\right)\mu=\frac{3}{4\pi G}\left(\frac{da}{adt}\right)\left[\left(\frac{da}{adt}\right)^2+\frac{k\Lambda}{2\pi^2}\right]}.
\end{array}
\end{equation}

{\bf Model III.}
In the presence of the fundamental constant $\mu_{GX}\equiv\frac{\Lambda}{8\pi G}$, with the variation of the gravitational constant $G=\frac{\pi}{4\mu_{GX}}\left(\frac{c}{a}\right)^2$ and $c=$const, cosmological equations appear as

\begin{equation}
\begin{array}{l}
\displaystyle{\left(\frac{da}{adt}\right)^2+\frac{k c^2}{a^2}=\frac{2\pi^2}{3}\left(\frac{c}{a}\right)^2\left(1+\frac{\mu}{\mu_{GX}}\right)}, \\
\displaystyle{\dot\mu+3\left(\frac{da}{adt}\right)\mu=2\left(\frac{da}{adt}\right)\mu\left(1+\frac{\mu_{GX}}{\mu}\right)}.
\end{array}
\end{equation}

{\bf Model IV.}
Finally, with constant vacuum energy density $\rho_{GX}\equiv\frac{\Lambda c^2}{8\pi G}$, varying speed of light $c=\left(\frac{4G\rho_{GX}}{\pi}\right)^{1/4}a^{1/2}$ and $G$=const we have
\begin{equation}
\begin{array}{l}
\displaystyle{\left(\frac{da}{adt}\right)^2=\frac{8\pi G}{3}\mu+\frac{\beta}{a}}, \\
\displaystyle{\dot\mu+3\left(\frac{da}{adt}\right)\mu=\frac{3}{8\pi G}\left(\frac{da}{adt}\right)\left[\left(\frac{da}{adt}\right)^2+\frac{\beta}{a}\,\frac{2\pi^2+3k}{2\pi^2-3k}\right]},
\end{array}
\end{equation}
where $\beta=\frac{4\pi^2}{3}\left(G\rho_{GX}\right)^{1/2} \left(1-\frac{3k}{2\pi^2}\right)$.

In order to analyze the corresponding dynamical systems at infinity we perform the following transformations
\begin{equation}
\begin{array}{l}
\displaystyle{\alpha=\frac{2}{\pi}\textrm{arctg}(a)}, \quad\quad\quad	\displaystyle{m=\frac{2}{\pi}\textrm{arctg}(\mu)},
\end{array}
\end{equation}
so that the limits $a\rightarrow\infty$ and $\mu\rightarrow\infty$ correspond to $\alpha\rightarrow 1$ and $m\rightarrow 1$.

\begin{figure}[htp]
	\centering
		\includegraphics[width=5in]{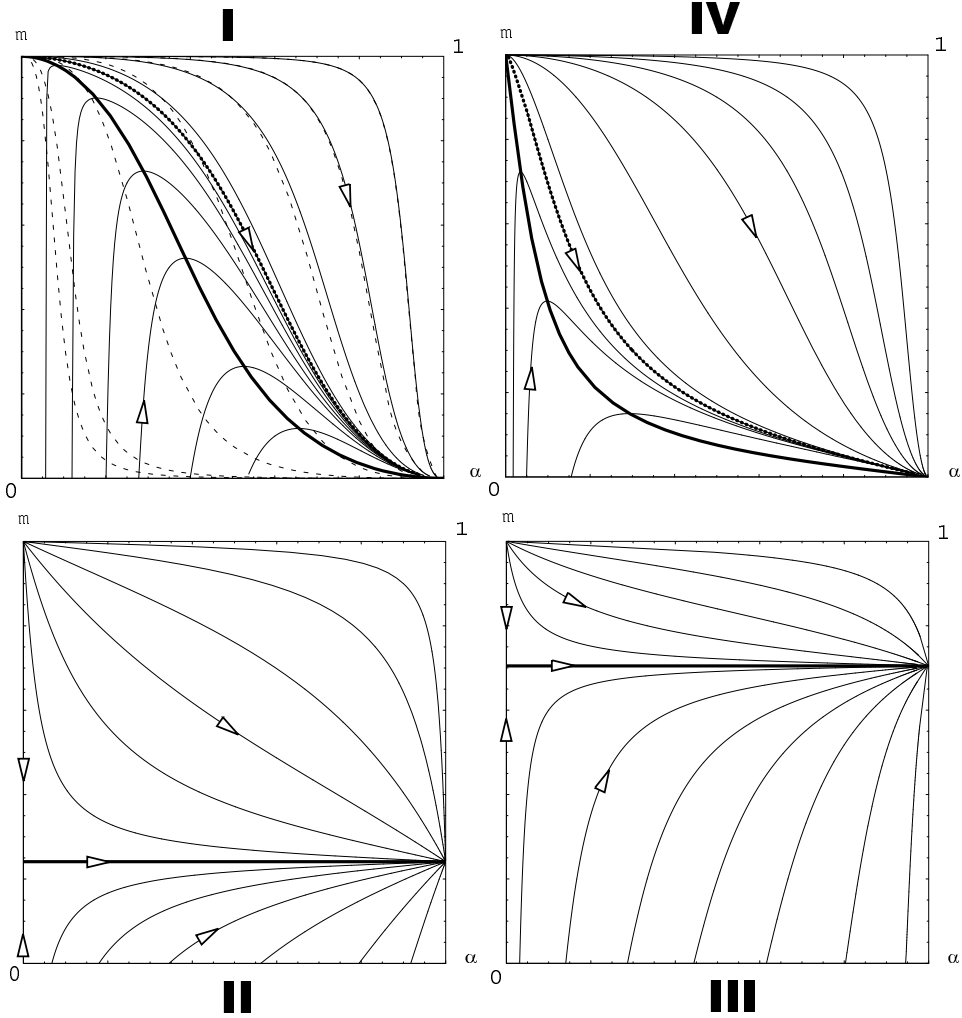}
	\caption{Phase portraits for models I-IV. Directions of phase trajectories are shown by arrows. See detailed explanations in the text.}
	\label{pp}
\end{figure}

The phase portraits for models I-IV are shown in fig.\ref{pp} in variables $\left\{m,\alpha\right\}$ for the case $k=0$ at the expansion phase $(da/dt>0)$.
The upper left corner in each diagram corresponds to the Friedmann cosmological singularity with $\mu=\infty$ and $a=0$. In the lower right corner $\mu\rightarrow 0$ and $a\rightarrow\infty$. For model I we also present Friedmannian usual solutions $\mu\propto a^{-3}$, which are shown by dashed curves in fig.\ref{pp}.

Two characteristic curves can be found for models I and IV: the \emph{separatrix}, shown by thick dotted curve, and the thick curve (given by equality $dm/d\alpha=0$) corresponding to \emph{maximum density} for each solution below the separatrix. Above the separatrix, the solutions are similar to the usual Friedmannian ones, namely, they start with a classical singularity and end up with infinite scale factor and zero density. In contrast, solutions below the separatrix start at a positive scale factor and zero density and tend to the same limit as above, so that each of them passes through maximum density, indicated by a thick curve.

For models II and III the separatrix is the horizontal line, shown by the thick curve. It also separates two regions in the phase portraits: above the separatrix the solutions start at the Friedmannian singularity, but end up in the de-Sitter phase with $\mu\rightarrow$const and $a\rightarrow\infty$; below the separatrix the solutions start, similarly to models I and IV, with zero density and positive scale factor, ending in the same de-Sitter limit.

We observe similarity of phase portraits corresponding to the pairs of models (I,IV) and (II,III), which is not evident from the cosmological equations. In fact, model II possesses analytical solution. The Hubble parameter $da/(adt)$ can be expressed from the first equation and substituted into the continuity equation, leading to the following solution
\begin{equation}
\mu(t)=\frac{\Lambda}{4\pi G}\left[1-\frac{\frac{3}{2}\left(1-\frac{k}{2\pi^2}\right)} {\cosh^{2}\left\{\frac{1}{2}\left[\Lambda\left(1-\frac{k}{2\pi^2}\right)\right]^\frac{1}{2}t +\textrm{arcsech}\left(\frac{2}{3}\,\frac{1-4\pi G\mu_0\Lambda^{-1}}{1-\frac{k}{2\pi^2}}\right)^\frac{1}{2}\right\}}\right],
\label{solII}
\end{equation}
where $\mu_0$ is the value of the matter density at $t=0$. In contrast, there is no analytical solution for model III.

Obviously, the precise position of each separatrix is determined by values of the corresponding parameters, such as $c$, $G$, $\mu_{GX}$ and $\beta$ which are set to unity everywhere, except for model II, where we took $\Lambda=5$ to shift the separatrix away from the lower border.

In principle, given the solution with \emph{actual} values of parameters, we have to find the location of the separatrix in the phase diagram. Each solution can be parametrized by only one quantity $\Omega_m\equiv\frac{8\pi G}{3H_0^2}\mu_0$, the present dimensionless matter density, and the fraction left for curvature (model I) and dark energy (models II-IV). The separatrix is given by the following condition\footnote{For models II and III this condition follows from $dm/d\alpha=0$.}
\begin{eqnarray}
	1-4\pi G\mu_0\Lambda^{-1}=0,
\label{sepII}
\end{eqnarray}
which leads to equation
\begin{eqnarray}
	1-\frac{\Omega_m}{2(1-\Omega_m)}=0,
\end{eqnarray}
having as a solution $\Omega_{sep}=2/3$. Surprisingly enough, we find that the value of this crucial parameter, which defines the position of the separatrix, is the same for all four models, even though they are given by quite different set of equations! In addition, this equation is valid for any spatial curvature.

This fact points out a remarkable underlying invariance possessed by the GX-models as a result of the scaling (\ref{Lambda}).

We are thankful to the referee for a number of helpful comments.


\begin{thebibliography}{xx}
\bibitem{GX}
V. G. Gurzadyan, S.-S. Xue in: ``From Integrable Models to Gauge Theories; volume in honor of Sergei Matinyan'', ed. V. G. Gurzadyan, A. G. Sedrakian, p.177, {\it World Scientific}, 2002; {\it Mod. Phys. Lett.} {\bf A18} (2003) 561 [astro-ph/0105245, see also astro-ph/0510459].
\bibitem{barrow}
T. Clifton, J. D. Barrow and R. J. Scherrer, {\it Phys.Rev.} {\bf D71} (2005) 123526; 
D. Parkinson, B. A. Bassett and J. D. Barrow, {\it Phys.Lett.} {\bf B578} (2004) 235.
\bibitem{Ver05} G.V. Vereshchagin, {\it Mod. Phys. Lett.} {\bf A21} (2006) 729 [astro-ph/0511131].
\bibitem{Ver06}
G. V. Vereshchagin, G. Yegorian, submitted to JCAP [astro-ph/0601073].
\bibitem{Bel85}
V. A. Belinsky, L. P. Grishchuk, I. M. Khalatnikov, and Y. B. Zeldovich, {\it Phys. Lett.} {\bf B155} (1985) 232.
\bibitem{DG}
S. G. Djorgovski, V. G. Gurzadyan, talk at 7th UCLA Symposium, Dark Matter - 2006 (Marina del Rey, Feb., 2006) http://www.astro.caltech.edu/~george/de/
\bibitem{P}
T. Pandmanabhan, {\it Class Quant.Grav.} {\bf 22} (2005) L107; astro-ph/0603114
\end{thebibliography}
\end{document}